# A Calculation of the Physical Mass of Sigma Meson

J. R. Morones Ibarra[1], and Ayax Santos Guevara[1]

(1) Facultad De Ciencias Físico-Matemáticas, Universidad Autónoma de Nuevo León, Ciudad Universitaria, San Nicolás de los Garza, Nuevo León, 66450, México.

**Abstract**

We calculate the physical mass and the width of the sigma meson by considering that it couples in vacuum to two virtual pions. The mass is calculated by using the spectral function, and we find that it is about 600$MeV$. In addition, we obtained the value of 250$MeV$ for the width of its spectral function. The value obtained for the mass, is in an excellent agreement with that reported in the Particle data Book for the $\sigma$-meson, which is also named $f_0(600)$. This result also shows that the sigma meson can be considered as a two pion resonance.

**Introduction**

Even though the existence of the scalar-isoscalar $\sigma$ meson is subject to controversy [1], the study of its properties and structure in vacuum and in nuclear matter has recently attracted growing attention because of the role it plays in some theoretical models and the increasing experimental evidence of its existence [2]. It has been assumed that the $\sigma$ meson participates as an intermediate particle in several processes in vacuum and in hot and dense nuclear matter [3]. Particularly, the sigma meson is of wide interest because in theoretical models of the nuclear force, the sigma is the responsible for the attractive part of the nuclear potential [4], where it is attributed to the exchange of a single $\sigma$-meson

In addition, the sigma meson plays the role of the chiral partner of the pion in the sigma model, which is a toy model for the interaction between nucleons and pions, with $SU(2)_L \otimes SU(2)_R$ symmetry [5,6,7,8]. This model was originally developed by M. Gell-Mann and M. Levy [9], being the σ- meson analogous to the Higgs particle in the Weinberg-Salam theory [7,9], in the sense that the nucleon gets mass when the $SU(2)_L \otimes SU(2)_R$ symmetry is spontaneously broken. Moreover in chiral perturbation theory the sigma meson enters as an essential part to adjust the theory to the experimental data [10]. In [6] they propose some experimental possibilities for investigating the behaviour of the sigma meson in hot and in dense nuclear matter.

On the other hand, it is still under discussion the existence and the composition of the $\sigma$-meson, as a $\bar{q}q$ meson or as a $\pi\pi$ resonance [11,12,13] . We conclude here that it can be considered as a two pion resonance.

There are several ways of defining and determining theoretically the mass and the width of unstable particles. For some authors, as in [14], the mass of a particle is defined as the pole in its complete propagator; this definition is used also in [15,16,17]. On the other hand, authors as in [18,19] make use of the spectral function to define a mass of a particle. In [20,21] the S-matrix formalism is used to determine the mass and the width of a meson.

In this work we define the physical mass and the width in terms of the spectral function. We study the spectral function of σ- meson in vacuum when this meson couples through the strong interaction to two virtual pions. Defining the physical mass of a particle as the magnitude $|k|$ of its four-momentum for which the particle spectral function $S(k^2)$ gets its maximum, we find a closed expression for the regularized self-energy function of the σ- meson and so we obtain an exact analytical function for its the physical mass. The σ- meson self- energy is calculated in the one loop level and the propagator is computed by summing over ring diagrams, in the so called Random Phase Approximation (RPA) [22], which is characterized by the calculation of the self-energy to one loop order. To carry out the summation we use the Dyson's equation. The real part of the self-energy is ultraviolet divergent and it is regularized by using a double subtracted dispersion relation, which preserves the symmetries of the theory.

**Formalism**

The spectral function of a particle is defined as $-2\pi$ times the imaginary part of its propagator. The technique of defining the physical mass of a particle in terms of its spectral function, is used extensively in the literature, [18,23,24] and it is well established. This method will be used here for calculating the physical $\sigma$-meson mass.

In order to evaluate the physical mass of the σ- meson we need, firstly, to calculate its dressed propagator $\Delta(k)$ in vacuum, where $k$ is the four-momentum of the propagating meson. The expression for the dressed σ- meson propagator is obtained from the Dyson equation, [4,25]

$$i\Delta(k) = i\Delta_0(k) + i\Delta_0(k)[-i\Sigma(k)]i\Delta(k) \tag{1}$$

where $$i\Delta_0(k) = \frac{i}{k^2 - (m_\sigma^0)^2 + i\varepsilon}$$

is the free σ- meson propagator, with $m_\sigma^0$ and $\Sigma(k)$ being the bare mass and the self-energy of σ, respectively. The self-energy $\Sigma(k)$ contains all the information about the interactions of the meson with the quantum vacuum, so, in order to determine the self energy $\Sigma(k)$ we must specify the dynamical content of our model.

Our starting point is the interaction Lagrangian density $L_{\sigma\pi\pi}$ which describes the σ-π dynamics, [26] given by

$$L_{\sigma\pi\pi} = \frac{1}{2}g_{\sigma\pi\pi}m_\pi \vec{\pi} \cdot \vec{\pi}\sigma \tag{2}$$

where $\vec{\pi} = (\pi_1, \pi_2, \pi_3)$ represents the Cartesian components of the pseudoscalar $\pi$-meson field, $\sigma$ is the scalar $\sigma$-meson field, $g_{\sigma\pi\pi}$ is the coupling constant, and $m_\pi$ is the mass of the $\pi$-meson.

The influence of the interaction of $\sigma$-mesons with virtual pions is introduced through the modification of the free propagator in the one loop approximation; this is shown graphically in the figure 1. This Feynman diagram contributes to the $\sigma$-meson self-energy. The dashed lines represent the $\sigma$-meson and the dotted lines represent the pion field. We will calculate the full propagator in the chain approximation, which consists in an infinite summation of the one loop self-energy diagrams [16]. The diagrammatical representation of the modified propagator is showed in figure 2, and the analytical expression is given by $i\Delta(k)$ in the Eq. 1

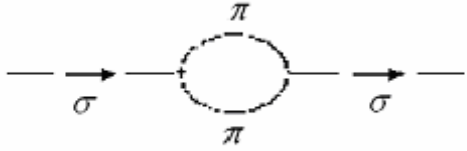

Figure 1. Sigma meson self-energy diagram. The dashed lines represent the $\sigma$-meson and the dotted lines the pion field

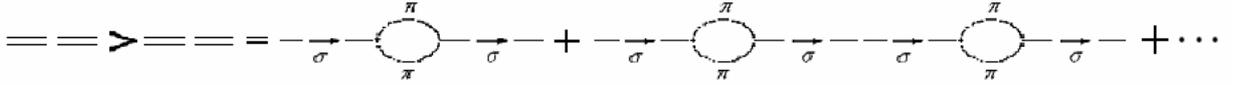

Figure 2. Sigma meson full propagator in the chain approximation. The double dashed line at the left represents the dressed sigma propagator and the right side is the diagrammatical representation of the chain approximation to the dressed propagator.

The solution for $\Delta(k)$ in the Dyson's equation (1) is given by

$$\Delta(k) = \frac{1}{[\Delta_0(k)]^{-1} - \Sigma(k)} = \frac{1}{k^2 - (m_\sigma^0)^2 - \Sigma(k)} \qquad (3)$$

On the other hand, the analytical expression for the self-energy $\Sigma(k)$, is given by [5,26]

$$-i\Sigma(k) = \frac{3}{2} g_{\sigma\pi\pi}^2 m_\pi^2 \int \frac{d^4q}{(2\pi)^4} \frac{1}{q^2 - m_\pi^2 + i\varepsilon} \frac{1}{(q-k)^2 - m_\pi^2 + i\varepsilon} \qquad (4)$$

where the coefficient 3/2 for the pion loop comes from the three isospin states and the permutation symmetry factor [26].

We will work in the σ-meson rest frame, for which $k^\mu = (k^0, \vec{0})$.

Carrying out the integration of $\Sigma(k)$ respect to $q_0$ by using the Cauchy residue theorem, and integrating in the $q_0$ complex plane, we obtain

$$\Sigma(k) = -\frac{3}{8\pi^2} g_{\sigma\pi\pi}^2 m_\pi^2 \int_{-\infty}^{\infty} \frac{\vec{q}^2 d|\vec{q}|}{\sqrt{(\vec{q}^2 + m_\pi^2)}\left[4(\vec{q}^2 + m_\pi^2) - k_0^2 - i\varepsilon\right]} \quad (5)$$

The real and imaginary parts of Eq. (5) can be separated with the use of the well known formula

$$\frac{1}{x \pm i\varepsilon} = P\frac{1}{x} \mp i\pi\delta(x) \quad (6)$$

The integration of the imaginary part is readily obtained, given the result

$$\mathrm{Im}\,\Sigma(k^2) = \frac{-3g_{\sigma\pi\pi}^2}{32\pi} m_\pi^2 \left(1 - \frac{4m_\pi^2}{k^2}\right)^{1/2} \quad (7)$$

for $k^2 \geq 4m_\pi^2$, and zero for $k^2 < 4m_\pi^2$. We can see the characteristic threshold value $k^2 \geq 4m_\pi^2$ for the production of real $\pi - \pi$ pairs from the σ-field.

On the other hand, the real part of $\Sigma(k^2)$ is ultraviolet divergent, and therefore it needs to be regularized. The regularization of $\mathrm{Re}\,\Sigma(k^2)$ will be done by using a double subtraction dispersion relation [27], which is given by the expression

$$\Sigma(t) = \frac{t^2}{\pi} \int_0^\infty \frac{\mathrm{Im}\,\Sigma(t')}{t'^2(t' - t) - i\varepsilon} dt' \quad (8)$$

For the imaginary part in the integrand taking from equation (7), the real part of the integral in Eq. (8) is convergent [27].

Now let us write the identity

$$\mathrm{Re}\,\Sigma(k^2) = \mathrm{Re}\,\Sigma(k^2) - \mathrm{Re}\,\Sigma_0(k^2) + \mathrm{Re}\,\Sigma_0(k^2) \quad (9)$$

where $\mathrm{Re}\,\Sigma_0(k^2)$ is an infinite quantity chosen conveniently to cancel the infinite terms of $\mathrm{Re}\,\Sigma(k^2)$. We define now the finite quantity $\mathrm{Re}\,\Sigma^R(k^2) = \mathrm{Re}\,\Sigma(k^2) - \mathrm{Re}\,\Sigma_0(k^2)$ as the regularized real part of the σ-meson self energy, and the renormalized mass $m_\sigma$ through $m_\sigma^2 = (m_\sigma^0)^2 + \mathrm{Re}\,\Sigma_0(k^2)$. $m_\sigma$ is actually the experimental mass for the sigma.

From Eq. (8), we separate the real and imaginary part by using again Eq. (6), obtaining for the real part, the expression

$$\mathrm{Re}\Sigma^R(k^2) = -\frac{3m_\pi^2 g_{\sigma\pi\pi}^2 k^2}{32\pi^2} P\int_{4m_\pi^2}^{\infty} \frac{(1-\frac{4m_\pi^2}{x'})^{\frac{1}{2}}}{x'^2(x'-k_0^2)} dx' \qquad (10)$$

This integral can be carried out directly giving the result

$$\mathrm{Re}\Sigma^R(k^2) = -\frac{3g_{\sigma\pi\pi}^2 k_0^2}{64\pi^2}\left(\frac{2}{3} - c + (c-c^2)I_0\right), \qquad (11)$$

where

$$c \equiv 1 - \frac{4m_\pi^2}{k^2} \text{ and,}$$

$$I_0 = \frac{1}{2\sqrt{c}}\ln\left|\frac{\sqrt{c}-1}{\sqrt{c}+1}\right| \text{ with } c > 0$$

The expression for the renormalized self-energy $\Sigma^R(k^2) = \mathrm{Re}\Sigma^R(k^2) + i\,\mathrm{Im}\Sigma(k^2)$, constructed from Eqs. (7) and (11), is the main result of this work.

The propagator given by Eq. (3), takes the form

$$\Delta(k^2) = \frac{1}{k_0^2 - m_\sigma^2 - \mathrm{Re}\Sigma^R(k^2) - i\,\mathrm{Im}\Sigma(k^2)} \qquad (12)$$

From the definition of the spectral function $S(k^2)$ given above, we have

$$S(k^2) = -\frac{2\pi\,\mathrm{Im}\Sigma(k^2)}{\left[k_0^2 - m_\sigma^2 - \mathrm{Re}\Sigma^R(k^2)\right]^2 + \left[\mathrm{Im}\Sigma(k^2)\right]^2} \qquad (13)$$

Substituting $\mathrm{Im}\Sigma(k^2)$ and $\mathrm{Re}\Sigma^R(k^2)$ from Eqs.(7) and (11) into Eq. (13), we obtain a closed expression for the spectral function. The parameters in Eq. (13) are the reported $m_\sigma$ value for the σ-meson mass, which we take as 500*MeV*, and the bare σππ coupling constant $g_{\sigma\pi\pi} = 12.8$ [26]. The plot of $S(k^2)$ given by Eq. (13) is shown in the figure 3.

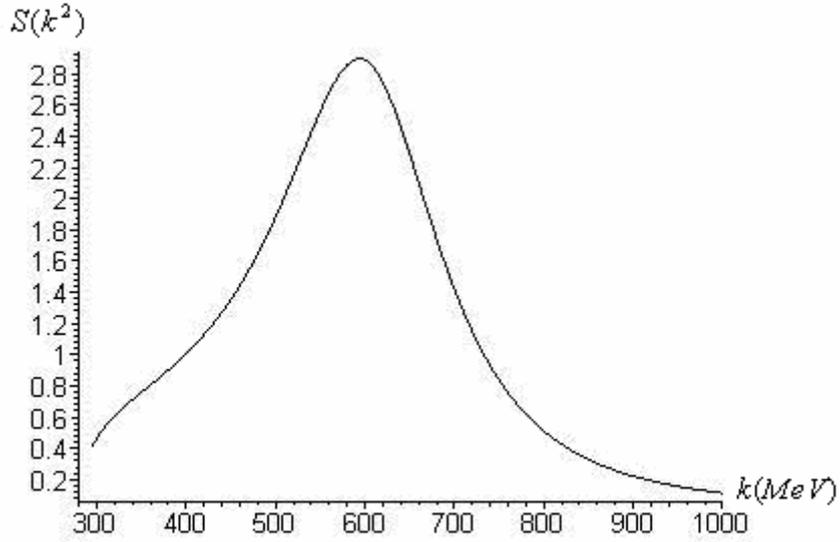

Figure 3. The σ-meson spectral function

As we can see, the maximum of $S(k^2)$ is getting at $|\vec{K}| = 600 MeV$, according to the reported value for the mass of σ [28]. On the other hand, we obtained the value of $250 MeV$ for the width, taking at one half of the maximum value of the spectral function.

**Conclusion**

We have succeeded in obtaining, in our calculations of the σ- meson physical mass, a value which is in fully agreement with that in the Particle data Book for $f_0(600)$, with a width for the spectral function of about $250 MeV$. The use of a double subtraction relation has allowed us to renormalize the σ- meson self-energy, given a closed analytical expression. The fact that we could obtained the physical mass of the σ- meson by considering that it couples in vacuum to two virtual pions, is a strong evidence that the σ- meson can be considered as a two pion resonance.